\documentclass[aps,pra,preprint,superscriptaddress,showpacs]{revtex4}
\begin{document}
\title{A genetic algorithm for finding pulse sequences for NMR quantum 
computing}
\author{Manoj Jesu Rethinam}
\affiliation{Department of Computer Science, \\ Wichita State 
University, 
Wichita, KS 67260-0083}
\author{Anil Kumar Javali}
\affiliation{Department of Computer Science, \\ Wichita State 
University, 
Wichita, KS 67260-0083}
\author{E.C.Behrman}
\email[]{elizabeth.behrman@wichita.edu}
\homepage[]{http://webs.wichita.edu/physics/behrman/behr.htm}
\affiliation{Department of Physics, \\ Wichita State University, 
Wichita, 
KS 67260-0032}
\author{J.E. Steck}
\affiliation{Department of Aerospace Engineering, Wichita State 
University, Wichita, KS 67260-0044}
\author{ S.R. Skinner}
\affiliation{Department of Electrical and Computer Engineering, Wichita 
State University, Wichita, KS 67260-0044}
\date{\today}

\begin{abstract}
We present a genetic algorithm for finding a set of pulse sequences, or 
rotations, for a given quantum logic gate, as implemented by NMR. We 
demonstrate the utility of the method by showing that shorter sequences 
than have been previously published can be found for both a CNOT and 
for the central part of Shor's algorithm (for N=15.) Artificial 
intelligence
techniques like the 
genetic algorithm here presented have an enormous potential for 
simplifying 
the implementation of working quantum computers.
\end{abstract}

\pacs{03.67.Lx,07.05.Mh}
\maketitle

\section{Introduction}
Of late the field of quantum computing has become of great interest, 
not least because of the possibilities of solving mathematical and 
physical problems either extremely difficult or impossible to compute 
on 
classical machines. There still remain, however, many obstacles to 
practical implementation. Some researchers have proposed the use of 
bulk 
samples at room temperature to do quantum computing: specifically, 
nuclear 
magnetic resonance (NMR) of molecules in a room temperature 
solution\cite{gc}\cite{chuang}\cite{jm}\cite{vander}.  Specific nuclear 
spins are the qubits. This approach has 
several advantages, including long coherence times for computation, 
ability to read out the state of each qubit spectroscopically, and 
ability 
to work at room temperature. In this implementation, the dynamics of 
the nuclear spins are controlled using RF pulses and delay times. 
However, 
complexity in designing a set of pulses grows rapidly with the size of 
the 
problem. The design of the pulse sequence for a given computation 
therefore becomes an important consideration: to find the shortest and 
most 
robust pulse sequence that implements the desired transformation. In 
this 
paper we propose a way of designing those pulse sequences, and 
demonstrate its utility by showing that we can increase the efficiency 
of the 
CNOT sequence by 40\% and of Shor's algorithm by 50\%.
\section{NMR}

Over several decades Nuclear Magnetic 
Resonance(NMR)\cite{nmr}\cite{chupp} has been 
widely used as a high sensitive technique for probing impurities in 
materials and for identifying molecules and determining their structure 
and dynamics. During the last few years\cite{jm}\cite{chuang},  the 
methods of NMR 
spectroscopy were applied to realize quantum computing experimentally. 
Each 
molecule is viewed as a single computer whose state is determined by 
the 
orientation of its spins.

Spin-1/2 nuclei are used for the purpose \cite{nc}. When such a spin is 
subjected to a strong static magnetic field $B_0$, it can align or 
anti-align with this static field, similar to a classical bar magnet. 
These 
two states are labeled as a logical 0 or 1. A spin precesses at a 
particular resonance frequency about this static field. Each spin can 
be 
individually addressed using RF pulses tuned to its resonance 
frequency, 
provided it belongs to a different type of nucleus or one with a 
different chemical environment\cite{nmr}. By timing the duration and 
phase of 
this pulse we can tip the spin into any direction. 

We can manipulate the state of each nucleus by applying a rotating RF 
magnetic field at its resonance frequency. These RF pulses and the 
coupling between the neighboring spins provide a mechanism to realize 
any two-input, one-output logic gate, building blocks for computation. 
Two-qubit gates 
require an interaction between the spins. The coherence times for 
spin-1/2 
nuclei in NMR is on the order of thousands of seconds\cite{gc}. The 
switching 
time for a two-qubit gate can be on the order of milli-seconds. Thus 
several tens, hundreds or thousands of operations can be performed 
using 
NMR.
\section{Genetic Algorithm}
The Genetic Algorithm is a model of machine learning which derives its 
behavior from a metaphor of the processes of evolution in 
nature\cite{garis}. This is done by the creation within a machine of a 
population 
of individuals represented by chromosomes, in essence a set of 
character 
strings that are analogous to the base-4 chromosomes that we see in our 
own DNA. The individuals in the population then go through a process of 
evolution\cite{genalg}\cite{mitchell}. This process leads to the 
evolution of 
populations of individuals that are better suited to their environment 
than the individuals that they were created from, just as in natural 
adaptation. The genetic algorithm is essentially a means of designing 
the 
environment and the evolution of the populations so that the desired 
characteristics are selected for.

At the beginning of the computation, an objective function is designed 
to evaluate how closely the individuals conform to the desired 
characteristic(s).  This function, the fitness function, is designed to 
give 
graded and continuous feedback about 
how well an individual conforms.  A number of individuals are randomly 
initialized.  The objective function is then evaluated for these 
individuals. The first generation is thus produced. Individuals are 
selected 
according to their fitness for the production of offspring.  In 
selection, the 
individuals producing offspring are chosen. Each individual in the 
selection pool receives a reproduction probability depending on their 
own 
objective value and the objective value of all other individuals in the 
selection pool. This fitness is used for the actual selection step 
afterwards.

Parents are recombined to produce offspring. Individuals with high 
fitness should have high probability of mating. Crossover represents a 
way 
of moving through the space of possible solutions based on the 
information gained from the existing solutions. All offspring will be 
mutated 
with a certain probability. The fitness of the offspring is then 
computed. The offspring are inserted into the population replacing the 
parents, 
producing a new generation. This cycle is performed until the 
optimization criteria are reached.
\section{Implementation}
Here we explain the actual functioning of the genetic algorithm to 
determine various pulse sequences which can be used to realize a CNOT 
gate, 
using 2 qubit NMR quantum computing. Previously, the CNOT gate has been 
realized 
using 
a single set of five pulse sequences\cite{gc}. The program 
which we have constructed, using genetic algorithm, finds an enormous 
number of pulse sequences in groups of four to realize a CNOT gate, 
which 
is a significant improvement over the already existing solution.

Initially, any genetic algorithm selects an initial population of the 
essential characteristics from a  set of values in the search space. 
The members of this population are called "chromosomes." In 
this case the characteristics are different angles and axes, since each 
pulse corresponds to a rotation\cite{gc}. As the possible range of 
angle
values for this application can be anything between 
$\angle$0 and 
$\angle$360, we define an upper bound and a lower bound for the search 
space, as $\angle$360 and $\angle$0 respectively. Nine bits are needed 
to represent each angle in this search space, providing a precision of 
one 
degree. We also have a set of axes(x,y,z) to choose from. Because of 
the 
group 
properties of rotations, the continuum of possible axes for any single 
rotation can be expressed as an ordered product of rotations about the 
three 
chosen (x,y,z) axes. The genetic algorithm therefore has to choose 
chromosomes which represent a 
sequence of rotations about three possible axes for each of two qubits 
and 
for each rotation angle. At first we show how we evolve new angles 
from the search space and then proceed further to explain the same 
technique with axes. 

Required mappings are done to the values to enhance the computational 
speed. The genetic algorithm randomly selects an angle $\theta$ and an 
axis {\sl i} from the respective search space for the angles and axes. 
Having done this, rotation about any axis $\alpha$, by an amount 
$\mid$ $\alpha$ $\mid$ is calculated from the following
formula\cite{baym},
\begin{equation}
R_{\vec{\alpha}}(\alpha) = \cos(\alpha/2)\cdot I - {\it i}
\vec{\sigma}\cdot 
\hat{\alpha}\sin(\alpha/2)        \label{eq:q1}
\end{equation}

where $\hat{\alpha}$ = ($\vec{\alpha}$ / $\mid$ $\vec{\alpha}$ $\mid$), I 
is the 
identity matrix and 
$\vec{\sigma}$=($\sigma_{x}$,$\sigma_{y}$,$\sigma_{z}$) 
is the Pauli operator.
As an example, let us consider the rotation of qubits A and B about the 
z-axis by an amount $\alpha=\mid$ $\vec{\alpha}$ $\mid$:
\begin{equation}
R_{zAB}(\alpha) = \cos(\alpha/2) \cdot {\it \hat{I}} + {\it i}
\sin(\alpha/2) S_{zAB}     \label{eq:q2}  
\end{equation}

where $S_{zAB}$ ={\it  $\sigma_{zA}$} $\otimes$ {\it $\sigma_{zB}$},
with $\otimes$ the outer product, and  {\it $\sigma_{z}$} is 
\begin{displaymath}
\left( \begin{array}{cc} 
-1 & 0 \\ 
0  & 1 
\end{array} \right )  
\end{displaymath}

where we implicitly label the rows and columns by $\mid$0$>$, 
$\mid$1$>$ and $<$0$\mid$, $<$1$\mid$.
The rotation matrices are calculated and multiplied for the result 
\begin{equation}
Result = R_{\vec{\alpha_{1}}}(\alpha_{1}) \cdot 
R_{\vec{\alpha_{2}}}(\alpha_{2}) \cdots 
R_{\vec{\alpha_{N}}}(\alpha_{N}) \label{eq:q3}
\end{equation}

The result is checked for its correctness which is the fitness 
function.
\begin{equation}
FF = Result - desired  \label{eq:q4}
\end{equation}
For example, the CNOT gate is defined by the set of inputs with 
corresponding outputs, given by:
\begin{table}[h]
\begin{center}
\begin{tabular}{|c|c|}
\hline
Inputs & Outputs\\
\hline
0 \ \  0 & 0 \ \ 0\\
0 \ \  1 & 0 \ \ 1\\
1 \ \  0 & 1 \ \ 1\\
1 \ \  1 & 1 \ \ 0\\
\hline
\end{tabular}
\caption{CNOT inputs and outputs.  This is the controlled NOT because 
the first bit is unchanged, but the second is flipped iff the first is 
a 
1.}
\end{center}
\end{table}

This transformation is represented by the matrix
 \begin{displaymath} 
\left( \begin{array}{cccc} 
1 & 0 & 0 & 0\\ 
0 & 1 & 0 & 0\\
0 & 0 & 0 & 1\\
0 & 0 & 1 & 0
\end{array} \right ) 
\end{displaymath}

where we implicitly label the rows and columns by the bit values of the 
two qubits: 
$\mid$00$>$,$\mid$01$>$,$\mid$10$>$,$\mid$11$>$ and 
$<$00$\mid$,$<$01$\mid$,$<$10$\mid$,$<$11$\mid$, each of which 
represents a possible input or output state. For example, the one in the third 
row of the matrix means that the input state $\mid$10$>$ is mapped onto 
the output state $<$11$\mid$.

Usually one does not get the desired output from the initial 
population. Hence the next step is to go for crossover. We select two 
chromosomes from the search space as the "parents." We have used single 
point crossover 
technique. A point for crossover is selected randomly. The first 
portion of the 
bits of the first parent is combined with the second portion of the 
bits to generate one new offspring and the first portion of the bits of 
the second parent is combined with the second portion of the bits of 
the 
first parent to generate the second new offspring. This operation is 
performed on all parents to produce new offspring.
\begin{figure}
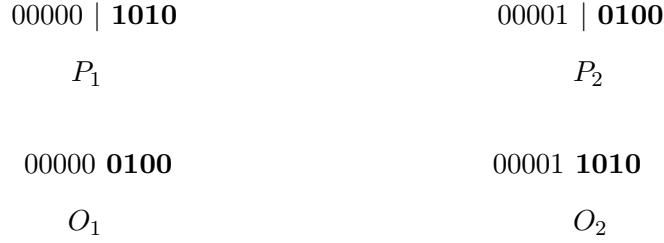

\begin{center}
$00000$ $\mid$ {\bf 1010} \hspace{4cm} $00001$ $\mid$ {\bf 0100}
\\ {\bf $P_{1}$} \hspace{6cm} {\bf $P_{2}$}
\end{center}
\begin{center} 
$00000$ {\bf 0100} \hspace{4cm} $00001$ {\bf 1010}
\hspace{2cm} \\ {\bf $O_{1}$} \hspace{6cm} {\bf $O_{2}$} 
\end{center}
\caption{An example of the crossover procedure for producing offspring 
$O_1$ and $O_2$ from parents $P_1$ and $P_2$}
\end{figure}
Figure 1 shows two parents ($P_1$,$P_2$) before crossover and the 
offspring($O_1$,$O_2$) after crossover. The point of crossover is 
randomly selected 
to be 5 in this example.  Having generated new offspring, we perform 
mutation to refine or fine tune the obtained new offspring, so that the 
resulting offspring 
will have better fitness. One of the chromosomes is affected after 
mutation. In other words, mutation acts on one chromosome at a time. 
This can 
either be an angle or an axis depending on what we perform the mutation 
on. Initially we select a value R[i], from the population. Two random 
numbers from the search space are selected and we perform a modulo 
operation with the mutation rate, selected earlier. If 
the 
result of the modulo function results in two identical numbers, then 
each bit of the value R[i] is flipped. The second value from the 
population is selected and the same process continues until all the 
values in 
the population have gone through this process of mutation.

We remap angles which exceed $\angle$360 to maintain a 9-bit length, by 
using the mapping function 
given 
below
\begin{equation}
r[i] = r[i] \times ((max-min)/(2^L-1))        \label{eq:q5}
\end{equation}
where r[i] is an array of integers representing the angles, max is 
the maximum value r[i] can hold from the search space, min is the 
minimum value r[i] can take from the search space, and L is the maximum 
number of bits 
used to represent the \{r[i]\}.
We have thus mapped the angles which exceed 360 to any value between 0 
and 360.

Having evolved a new set of values, rounding of the angles is performed 
in the same manner as it was done for the initial population. Now that 
we have found new angles we repeat the same procedure for the axes. In 
this case, the axes are allowed to crossover and the new generation 
just obtained is made to undergo mutation. Mutation can be defined as a 
method to fine tune the results obtained as a result of crossover. With 
these initial values derived, the fitness function is calculated as 
shown above in equation (3), and then the result is checked for 
correctness 
using equation (4). If the result is a zero, then the pulse sequence we 
have found can 
be used to realize a CNOT gate. If we get any other value, then the 
entire process of crossover, mutation and rounding of the angles 
followed by 
the crossover and mutation of the axes and evaluating the fitness 
function is performed. This 
process continues until a correct result is found.
\section{Results}
\subsection{CNOT}
We have realized a CNOT gate using pulse sequences with just 3
rotations thus increasing the efficiency by 40\% from the earlier
techniques 
which used 5 rotations. The two solutions found are given in Table 2.
\begin{table}[h]
\begin{center}
\begin{tabular}{|l|l|l|l|}
\hline
1 & $R_{zB}$(270) & $R_{xzAB}$(90) & $R_{xA}$(90)  \\
\hline
2 & $R_{xA}$(90) & $R_{zB}$(270) &  $R_{xzAB}$(90) \\
\hline
\end{tabular}
\caption{\label{tab:2/tc}Sample pulse sequences found by the genetic
algorithm, with 3 rotations, whose ordered product produces the CNOT
matrix.  Each pulse corresponds to a rotation  
$R_{\vec{\alpha}}(\alpha)$.}
\end{center}
\end{table}
\subsection{Shor's Algorithm}
Since its discovery \cite{shor} by Peter Shor in 1994, this algorithm 
for factorizing large numbers using a quantum computer has generated a 
large amount of interest\cite{ej}, not least because of its potential 
to 
render ineffective the RSA algorithm.  Chuang and co-workers 
\cite{chuang}\cite{vander} have shown a 
way to implement this algorithm for N=15 with the NMR system: a 
sequence of
steps whose core is two successive CNOT gates, $CNOT_{AB}$ and 
$CNOT_{AC}$.
As our genetic algorithm has increased the efficiency of the 
implementation of CNOT from 5 rotations to 3, the efficiency of Shor's
has 
increased from 10 to 6. But can we do still better? We decided to
simulate 
the product of the CNOT gates, which is the matrix \begin{displaymath} 
\left( \begin{array}{cccccccc} 
1 & 0 & 0 & 0 & 0 & 0 & 0 & 0\\ 
0 & 1 & 0 & 0 & 0 & 0 & 0 & 0\\
0 & 0 & 1 & 0 & 0 & 0 & 0 & 0\\
0 & 0 & 0 & 1 & 0 & 0 & 0 & 0\\
0 & 0 & 0 & 0 & 0 & 0 & 0 & 1\\
0 & 0 & 0 & 0 & 0 & 0 & 1 & 0\\
0 & 0 & 0 & 0 & 0 & 1 & 0 & 0\\
0 & 0 & 0 & 0 & 1 & 0 & 0 & 0
\end{array} \right ) 
\end{displaymath}

Where we implicitly label the rows and columns again by the values of 
bits A, B, and C, i.e, $\mid$000$>$, 
$\mid$001$>$, $\mid$010$>$, $\mid$011$>$, $\mid$100$>$, $\mid$101$>$, 
$\mid$110$>$, $\mid$111$>$ and $<$000$\mid$, $<$001$\mid$, 
$<$010$\mid$, 
$<$011$\mid$, $<$100$\mid$, $<$101$\mid$, $<$110$\mid$, 
$<$111$\mid$.The 
program found two solutions of only 5 rotations, which is an
improvement of 50\% . They are listed in Table 3.
\begin{table}[h]
\begin{center}
\begin{tabular}{|l|l|l|l|l|l|}
\hline
1 & $R_{xA}(90)$ & $R_{xC}$(250) & $R_{xCA}$(160) &
$R_{zxBC}$(90) & $R_{xzAB}$(260) \\
\hline
2 & $R_{xzAB}(80)$ & $R_{xA}(80)$ & $R_{zxBC}$(110) &
 $R_{xC}$(105) & $R_{zB}$(190) \\
\hline
\end{tabular}
\caption{\label{tab:3/tc}Sample pulse sequences found by the genetic
algorithm, with 5 rotations, whose ordered product produces the product
matrix $CNOT_{AB}$ $CNOT_{AC}$, the core calculation of Shor's
algorithm for N=15.
Each pulse corresponds to a rotation $R_{\vec{\alpha}}(\alpha)$.}
\end{center}
\end{table}
\section{Conclusions and Future Directions}
We have devised a new algorithm to genetically determine the pulse 
sequence necessary to realize a given quantum gate.  Our program 
largely 
simplifies the design of pulses and realizes a CNOT gate using a pulse 
sequence with just three rotations. This is an improvement in 
efficiency
of 40\%.  The
efficiency of the Shor's algorithm implementation has been improved by 
50\% for the N=15 case.

It should be noted that while the pulse sequences found are shorter, 
they may not necessarily be easier to implement. Several of the sequences 
found require interactions of types, like xz, which may not be easily 
accessible experimentally.  This would depend on the specific spin 
system being used.

It should also be noted that the product of the two CNOT gates forms 
the core of Shor's algorithm only for the N=15 case. For a different 
factorization a completely different sequence of CNOT and Toffoli gates may 
be necessary; however, it only underscores the usefulness of an AI 
approach, that one would need to find different sequencesof gates and 
therefore different sequences of pulses for each factorization calculation.

In practice, NMR quantum computing is more than simply designing a 
pulse sequence off line and then implementing it exactly as simulated. 
Any 
experimental setup inevitably contains factors and effects which no 
simulator, simplified as it must be, can take into consideration. 
Artificial intelligence
 techniques like the genetic algorithm are potentially powerful 
tools for the simplification of experimental quantum computing. The 
scientist constantly must twiddle with variables like the pulse widths 
to 
tune the system to the desired output; in principle, this could be done 
online with a genetic algorithm or neural network taking the 
scientist's place to find the 
optimal values for the variables.
\begin{acknowledgments}
This work was supported by National Science Foundation, 
Grant No.  ECS-0201995.We also acknowledge the Wichita State University 
High Performance 
Computing Center, and helpful discussions with 
Terry Miller, Mathias Steffen, L.M.K. Vandersypen, Sattiraju V.
Prabhakar, 
Tom Wallis, John Wang and Stuart Boehmer.
\end{acknowledgments}

\end{document}